\documentclass[aps,prc,twocolumn,showpacs,superscriptaddress,amsmath,amssymb]{revtex4}
\usepackage{bm}
\usepackage{dcolumn}
\usepackage{indentfirst}
\usepackage{longtable}
\usepackage{graphicx,epsfig,latexsym,amssymb}
\usepackage{multirow,amsmath,array,booktabs,color}
\usepackage[section]{placeins}

\newcommand{\beq}{\begin{equation}}
\newcommand{\eeq}{\end{equation}}
\newcommand{\beqn}{\begin{eqnarray}}
\newcommand{\eeqn}{\end{eqnarray}}
\newcommand{\bea}{\begin{array}}
\newcommand{\eea}{\end{array}}
\newcommand{\bsub}{\begin{subequations}}
\newcommand{\esub}{\end{subequations}}

\begin{document}

\title{Isospin coupling-channel decomposition of nuclear symmetry energy in covariant density functional theory}

\author{Qian Zhao}
\affiliation{School of Nuclear Science and Technology, Lanzhou University, Lanzhou 730000}
\affiliation{Key Laboratory of Special Function Materials and Structure Design, Ministry of Education, Lanzhou 730000}
\author{Bao Yuan Sun \footnote{Email: sunby@lzu.edu.cn}}
\affiliation{School of Nuclear Science and Technology, Lanzhou University, Lanzhou 730000}
\affiliation{Key Laboratory of Special Function Materials and Structure Design, Ministry of Education, Lanzhou 730000}
\author{Wen Hui Long}
\affiliation{School of Nuclear Science and Technology, Lanzhou University, Lanzhou 730000}
\affiliation{Key Laboratory of Special Function Materials and Structure Design, Ministry of Education, Lanzhou 730000}

\begin{abstract}
The isospin coupling-channel decomposition of the potential energy density functional is carried out within the covariant density functional (CDF) theory, and their isospin and density dependence in particular the influence on the symmetry energy is studied. It is found that both isospin-singlet and isospin-triplet components of the potential energy play the dominant role in deciding the symmetry energy, especially when the Fock diagram is introduced. The results illustrate a quite different mechanism to the origin of the symmetry energy from the microscopic Brueckner-Hartree-Fock theory, and demonstrate the importance of the Fork diagram in the CDF theory, especially from the isoscalar mesons, in the isospin properties of the in-medium nuclear force at high density.
\end{abstract}
\pacs{
21.60.Jz,  
21.65.Cd,  
21.65.Ef,  
21.65.Jk,  
21.30.Fe   
}

\maketitle

\section{Introduction}
Nuclear symmetry energy $E_S$, as defined by the difference of the binding energy per nucleon $E_b$ between symmetric nuclear matter (SNM) and pure neutron matter (PNM), plays an essential role in understanding the isospin-dependent aspects in nuclear physics and the critical issues in astrophysics, such as neutron skin thickness, dipole excitation modes of stable or exotic nuclei, as well as radius and cooling mechanism of neutron stars \cite{BA85,CJ86,BY78,WH85,M102,A107,BL464, WG779}.

Various of nuclear models have been attributed to the exploration of the symmetry energy. Among the phenomenological approaches, two typical versions of the density functional theory, i.e., non-relativistic Skyrme-Hartree-Fock (SHF) \cite{QP28} and relativistic mean field (RMF) \cite{BJ16} theories, are widely adopted to study the issues related to symmetry energy. Recently, the density-dependent behavior of the symmetry energy has been analyzed systematically in SHF theory with 21 sets of Skyrme interactions \cite{LW72} and in RMF theory with 23 sets of nonlinear, density-dependent or point-coupling interactions \cite{LW76}. It was found that remarkable divergence of $E_S$ appears when the density of nuclear matter increases beyond the saturation.

Furthermore, considering the Fock terms in the meson-exchange diagram, the density-dependent relativistic Hartree-Fock (RHF) theory \cite{WN640} has been developed. Both RMF and RHF theory are usually called as covariant density functional (CDF) theory. It has been found that the inclusion of the Fock terms strongly affect the density-dependent behaviors of symmetry energy at high densities, and in turn the radius and cooling process of neutron star \cite{BY78,WH85}. In particular, significant contributions to the symmetry energy have been found from the Fock terms in the isoscalar $\sigma$ and $\omega$ couplings, and the neutron-star properties determined by density-dependent RHF theory were shown to be in fairly good agreement with the observational data. Besides, the density dependence of symmetry energy is also described by microscopic calculations, such as variational approach \cite{A58}, Brueckner-Hartree-Fock (BHF) approach \cite{E72,ZH78,G87} and chiral effective field theory \cite{LL89}. These calculations predict a relatively narrow band compared to the data extracted from the isobaric analog states experiments below the saturation density, but divergence emerges at higher densities.

In principle, the symmetry energy characterizes the isospin structure of the in-medium nucleon-nucleon ($NN$) interactions, i.e., $nn$ and $pp$ interactions versus $np$ interaction \cite{AW411}. In view of the lack of enough knowledge about the isospin-dependence of in-medium nuclear effective interactions, the investigation on the symmetry energy, especially its behavior at high densities, from variety of theoretical models in collaboration with the constraints from the experiments such as heavy-ion collisions \cite{BL464}, could provide us a novel probe about the isospin nature of nuclear force.

The isospin structure of the in-medium $NN$ interaction can be described by decomposing the interacting two-body potentials according to two-nucleon isospin-states, i.e., isospin-singlet (T=0) potentials and isospin-triplet (T=1) potentials. In the microscopic BHF approach, T=0 and T=1 components can be obtained by the partial wave decomposition of the potential energy. Thus, the contributions from the different spin-isospin channels to $E_S$ and its slop parameter $L$ could be analyzed quantitatively \cite{IA84}. In CDF theory, the isospin structure of the in-medium $NN$ interactions is naturally taken into account via the meson-exchange mechanism. However, the detailed analysis of the isospin coupling-channel decompositions of the potential energy and their contributions to the symmetry energy has not been carried out yet in CDF framework. Particularly, in the density-dependent RHF theory, how the Fock diagram affects the isospin structure of the in-medium $NN$ interactions need to be clarified. Therefore, in this work we will study the isospin dependence of in-medium $NN$ interaction based on several selected CDF energy functionals and their influence on the properties of symmetry energy.

\section{Isospin coupling-channel decompositions in CDF energy functional}
The detailed description of the CDF theory can be found in Ref.\cite{WH69,BY78}. In this section we briefly layout the main CDF formulism in nuclear matter, and present the isospin coupling-channel decompositions of the potential energy density in terms of the two-nucleon isospin-states. In CDF theory, the energy functional is obtained by taking the expectation of the Hamiltonian with respect to the Hartree-Fock ground state, which consists of three parts
\begin{eqnarray}
\varepsilon_k&=&\sum_{ps\tau}\bar{u}(p,s,\tau)(\bm{\gamma}\cdot\bm{p}+M)u(p,s,\tau),\\
\varepsilon_\phi^D&=&\frac{1}{2}\sum_{p_1s_1\tau_1}\sum_{p_2s_2\tau_2}\bar{u}(p_1,s_1,\tau_1)\bar{u}(p_2,s_2,\tau_2)
\nonumber\\
&\times&\frac{\Gamma_\phi(1,2)}{m_{\phi}^2}u(p_2,s_2,\tau_2)u(p_1,s_1,\tau_1),\label{ED}\\
\varepsilon_\phi^E&=&-\frac{1}{2}\sum_{p_1s_1\tau_1}\sum_{p_2s_2\tau_2}\bar{u}(p_1,s_1,\tau_1)\bar{u}(p_2,s_2,\tau_2)\nonumber\\
&\times&\frac{\Gamma_\phi(1,2)}{m_{\phi}^2+\bm{q}^2}u(p_1,s_1,\tau_1)u(p_2,s_2,\tau_2),
\end{eqnarray}
with $\phi=\sigma,\omega,\rho,\pi$. Here $\varepsilon_k$ denotes the kinetic energy density, and $\varepsilon_\phi^D$ and $\varepsilon_\phi^E$ correspond to the Hartree (direct) and Fock (exchange) terms of the potential energy density which deduced from the two-nucleon interaction parts of the Hamiltonian. $\Gamma_\phi(1,2)$ represents various meson-$NN$ interacting vertex. Dirac spinors $u(p,s,\tau)$ with coordinate of momentum $p$, spin $s$ and isospin $\tau$ read as
 \begin{equation}\label{positive energy solutions}
    u(p,s,\tau)=\left[\frac{E^*+M^*}{2E^*}\right]^{1/2}
        \left(\begin{array}{c}
            1 \\ \displaystyle\frac{\boldsymbol{\sigma}\cdot\boldsymbol{p}^*}{E^*+M^*}
        \end{array}\right)
        \chi_s \chi_\tau,
 \end{equation}
for the positive energy, where $\chi_s$ denotes the spin wave function, and $\chi_\tau$ is the isospin wave function. In the following discussion we take $\tau=1/2$ for neutron, and $\tau=-1/2$ for proton. Then we can define two-nucleon isospin-singlet state $|00\rangle$ and isospin-triplet states $|11\rangle, |10\rangle, |1-1\rangle$,
\begin{eqnarray}
|00\rangle&=&\frac{1}{\sqrt{2}}\Big\{\chi_{n}(1)\chi_{p}(2)-\chi_{p}(1)\chi_{n}(2)\Big\},\\
|11\rangle&=&\chi_{n}(1)\chi_{n}(2),\\
|10\rangle&=&\frac{1}{\sqrt{2}}\Big\{\chi_{n}(1)\chi_{p}(2)+\chi_{p}(1)\chi_{n}(2)\Big\},\\
|1-1\rangle&=&\chi_{p}(1)\chi_{p}(2).
\end{eqnarray}

With these isospin-states the Hartree (direct) term of the potential energy density Eq.~(\ref{ED}) is divided into the isospin-singlet (T=0) and isospin-triplet (T=1) parts
\begin{eqnarray}
\varepsilon^{D}_\phi=\varepsilon^{D}_{\phi,T=0}+\varepsilon^{D}_{\phi,T=1},
\end{eqnarray}
where
\begin{eqnarray}
\varepsilon^{D}_{\phi,T=0}&=&\frac{1}{2}\bar{u}(p_1,s_1)\bar{u}(p_2,s_2)\Big\langle00\Big|\frac{\Gamma_\phi(1,2)}{m_{\phi}^2}\Big|00\Big\rangle\nonumber\\
&\times&u(p_2,s_2)u(p_1,s_1),
\end{eqnarray}
and
\begin{eqnarray}
\varepsilon^{D}_{\phi,T=1}&=&\frac{1}{2}\bar{u}(p_1,s_1)\bar{u}(p_2,s_2)\Big\{\Big\langle11\Big|\frac{\Gamma_\phi(1,2)}{m_{\phi}^2}\Big|11\Big\rangle
\nonumber\\
&+&\Big\langle10\Big|\frac{\Gamma_\phi(1,2)}{m_{\phi}^2}\Big|10\Big\rangle
+\Big\langle1-1\Big|\frac{\Gamma_\phi(1,2)}{m_{\phi}^2}\Big|1-1\Big\rangle\Big\}\nonumber\\
&\times&u(p_2,s_2)u(p_1,s_1).
\end{eqnarray}
Following the similar procedure the corresponding parts from the Fock term $\varepsilon^{E}_{\phi,T=0}$ and $\varepsilon^{E}_{\phi,T=1}$ are also obtained, which are not listed here. Correspondingly, the binding energy per nucleon in nuclear matter at a given baryon density $\rho_b$ and isospin asymmetry parameter $\delta=(\rho_n-\rho_p)/\rho_b$ can be separated into
\begin{eqnarray}
E_b(\rho_b,\delta)&=&E_{b,k}+E_{b,T=0}^{D}+E_{b,T=0}^{E}\nonumber\\
&+&E_{b,T=1}^{D}+E_{b,T=1}^{E},\label{Eb}
\end{eqnarray}
where $E_{b,k}$ is the contributions from the kinetic energy. $E_{b,T=0}^{D}$ and $E_{b,T=0}^{E}$ are Hartree and Fock terms in T=0 part of the potential energy, respectively, while $E_{b,T=1}^{D}$ and $E_{b,T=1}^{E}$ are corresponding T=1 parts.

In general, the energy per nucleon of asymmetric nuclear matter $E_b(\rho_b,\delta)$ can be expanded in the Taylor series with respect to the asymmetry parameter $\delta$, and the
density-dependent symmetry energy is deduced from the second order coefficient
\begin{eqnarray}
E_S(\rho_b)=\frac{1}{2}\frac{\partial^2E_b(\rho_b,\delta)}{\partial\delta^2}\Big|_{\delta=0},
\end{eqnarray}
which represents the isospin-dependent part in the energy functional of asymmetric nuclear matter. To shed light on the density dependence of the symmetry energy, $E_S$ can be further expanded around the nuclear saturation density $\rho_0$ as
\begin{eqnarray}
E_S(\rho_b)=~J+L\xi+\mathcal{O}(\xi^2),
\end{eqnarray}
where $J=E_S(\rho_0)$ and $\xi=(\rho_b-\rho_0)/3\rho_0$ is a dimensionless variable and
\begin{equation}
L=3\rho_0\frac{\partial E_S}{\partial\rho_b}\Big|_{\rho_b=\rho_0},
\end{equation}
is the slop parameter of symmetry energy. With the decompositions in Eq. (\ref{Eb}), we also have
\begin{eqnarray}
E_S(\rho_b)&=&E_{S,k}+E_{S,T=0}^{D}+E_{S,T=0}^{E}\nonumber\\
&+&E_{S,T=1}^{D}+E_{S,T=1}^{E},\\
L&=&L_{k}+L_{T=0}^{D}+L_{T=0}^{E}+L_{T=1}^{D}+L_{T=1}^{E}.
\end{eqnarray}
In this work we will study the density-dependent properties of symmetry energy based on several selected CDF energy functional, especially from a viewpoint of the isospin coupling-channel decomposition.

\section{Results and discussion}

Our calculations are based on the density-dependent effective interactions PKO1 \cite{WN640} and PKA1 \cite{WH76} within RHF theory, in comparison with TW99 \cite{ST656} and PKDD \cite{WH69} within RMF theory, since they have been utilized to study the bulk properties of asymmetric nuclear matter and neutron stars and reveal significant difference in describing the symmetry energy \cite{BY78,WH85}.

\subsection{Density dependence of symmetry energy}

\begin{figure}
  \centering
  \includegraphics[width=0.48\textwidth]{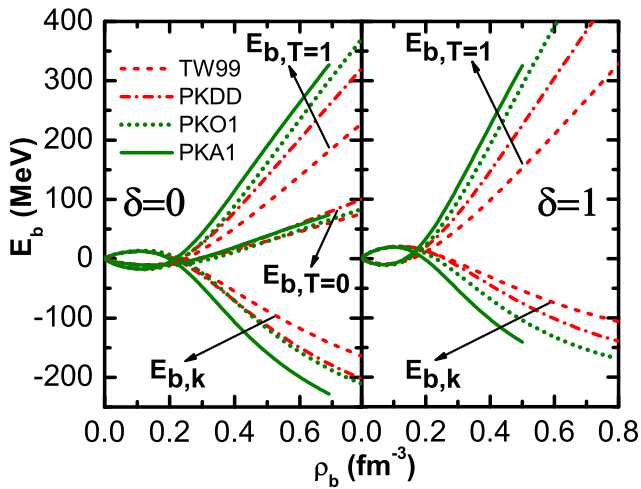}
  \caption{(Color online) The binding energy per nucleon $E_b$ is decomposed as kinetic energy part $E_{b,k}$, isospin-singlet potential part $E_{b,T=0}$ and isospin-triplet potential part $E_b^{b,T=1}$ in symmetric nuclear matter ($\delta=0$, left panel) and pure neutron matter ($\delta=1$, right panel) as function of baryon density $\rho_b$. The results are calculated by RHF effective interactions PKO1 and PKA1, in comparison with RMF ones TW99 and PKDD.}
\end{figure}

In Fig. 1, the decompositions of the binding energy per nucleon $E_b$, namely, the kinetic energy part $E_{b,k}$, isospin-singlet and isospin-triplet potential energy parts $E_{b,T=0}$ and $E_{b,T=1}$ are plotted in SNM ($\delta=0$) and PNM ($\delta=1$) with four selected CDF effective interactions. When the baryon density is smaller than about $0.2$ fm$^{-3}$, all of effective interactions predict nearly the same density-dependent behavior of the different components of $E_b$, while the divergence becomes larger with increasing density. It is found that at high density the contribution from the kinetic energy and T=1 potential parts is enhanced in PNM compared with those in SNM. Conversely, the T=0 potential energy parts $E_{b,T=0}$, which is attributed to the $np$ interaction of nuclear force, disappears in PNM due to the asymmetry between neutron and proton. Therefore, it is expected that distinct disparity of isospin dependence of these components will lead to the uncertainty in the symmetry energy.

In terms of isospin coupling-channel decomposition, we then obtain three components of nuclear symmetry energy, i.e., kinetic contribution $E_{S,k}$, isospin-singlet potential contribution $E_{S,T=0}$ ($=E_{S,T=0}^D+E_{S,T=0}^E$) and isospin-triplet potential contribution $E_{S,T=1}$ ($=E_{S,T=1}^D+E_{S,T=1}^E$), as shown in Fig. 2. One should notice that the contribution to $E_S$ from Hartree term in $\omega$-meson coupling-channel is vanished because of its isoscalar and Lorentz vector nature \cite{BY78}. The symmetry energy deduced from T=0 and T=1 components of Hartree term in $\omega$-meson coupling-channel compensate each other actually, i.e., $E_{S,T=0}^{D,\omega}+E_{S,T=1}^{D,\omega}=0$. Hence, in the following discussion we do not include these parts any more for convenience.

From Fig. 2, it is revealed that remarkable deviation between the RMF and RHF calculations arises even at low-density region. For $E_{S,k}$ and $E_{S,T=0}$, RMF effective interactions give larger values than RHF ones, while opposite conclusion occurs for $E_{S,T=1}$. Among three components, $E_{S,T=1}$ displays the most distinct deviation between the RMF and RHF calculations, which results in a stiffer symmetry energy in RHF at high density as has been illustrated in Ref. \cite{BY78}. In addition, it is seen in RMF cases $E_{S,k}$ and $E_{S,T=0}$ dominate the symmetry energy, while all of three components represent non-negligible contribution in RHF cases. Analysis from a point of view of in-medium $NN$ interactions could be useful to clarify these results.

\begin{figure}
  \centering
  \includegraphics[width=0.48\textwidth]{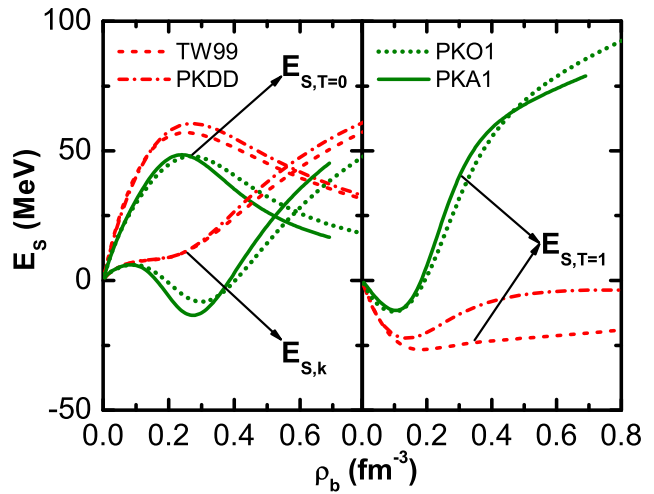}
  \caption{(Color online) The symmetry energy is decomposed as kinetic part $E_{S,k}$, isospin-singlet potential part $E_{S,T=0}$ (left panel) and isospin-triplet potential part $E_{S,T=1}$ (right panel) as function of baryon density $\rho_b$. The results are calculated by RHF effective interactions PKO1 and PKA1, in comparison with RMF ones TW99 and PKDD.}
\end{figure}

\begin{figure}
  \centering
  \includegraphics[width=0.48\textwidth]{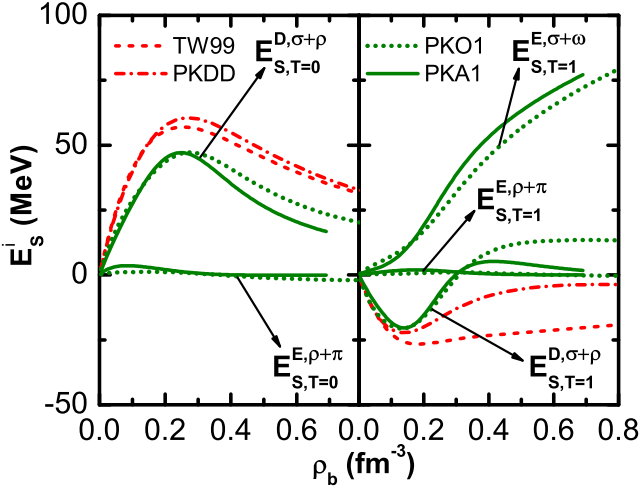}
  \caption{(Color online) The isospin-singlet potential symmetry energy $E_{S,T=0}$ is decomposed as the Hartree term's part $E_{S,T=0}^{D,\sigma+\rho}$ within $\sigma$ and $\rho$-meson coupling channels, and Fock term's part $E_{S,T=0}^{E,\rho+\pi}$ within $\rho$ and $\pi$-meson coupling channels as function of baryon density $\rho_b$ (left panel), while the isospin-triplet potential symmetry energy $E_{S,T=1}$ is divided into the Hartree term's part $E_{S,T=1}^{D,\sigma+\rho}$ within $\sigma$ and $\rho$-meson coupling channels, Fock term's part $E_{S,T=1}^{E,\rho+\pi}$ within $\rho$ and $\pi$-meson coupling channels, as well as Fock term's part $E_{S,T=1}^{E,\sigma+\omega}$ within $\sigma$ and $\omega$-meson coupling channels (right panel). The results are calculated by RHF effective interactions PKO1 and PKA1, in comparison with RMF ones TW99 and PKDD.}
\end{figure}

In Fig. 3, two isospin coupling-channel's components of the symmetry energy $E_{S,T=0}$ and $E_{S,T=1}$ are decomposed in terms of the direct and exchange terms in $NN$ interactions. For the cases of direct term, the results are attributed to the $\sigma$- and $\rho$-meson coupling-channels. The divergence between RMF and RHF takes place in both T=0 and T=1 components (see $E_{S,T=0}^{D,\sigma+\rho}$ in the left panel and $E_{S,T=1}^{D,\sigma+\rho}$ in the right panel of Fig. 3), which could be explained by the difference of the meson-nucleon coupling constants among the selected effective interactions. It is interesting to see that when summing up the contribution of T=0 and T=1 channels, namely, $E_{S,T=0}^{D,\sigma+\rho}+E_{S,T=1}^{D,\sigma+\rho}$, the total contribution from the direct term in $NN$ interactions is almost same for all of four CDF effective interactions. In fact, the deviation of $E_S$ between the RMF and RHF calculations results mainly from the exchange term in $NN$ interactions \cite{BY78}. To clarify this, in Fig. 3 the Fock term's parts of T=0 and T=1 symmetry energy are divided further according to the isoscalar and isovector meson-nucleon coupling-channels. We see that the contributions from the isovector coupling-channels, i.e., $E_{S,T=0}^{E,\rho+\pi}$ and $E_{S,T=1}^{E,\rho+\pi}$, are negligible in both $E_{S,T=0}$ and $E_{S,T=1}$ cases due to their small meson-nucleon coupling constants. However, the corresponding isoscalar coupling-channels' $E_{S,T=1}^{E,\sigma+\omega}$ exhibits a strong density-dependent behavior, which in turn leads to the distinct deviation of $E_{S,T=1}$ between the RMF and RHF calculations in Fig. 2. In fact, via the Fock diagram, the isoscalar $\sigma$- and $\omega$-meson account for only $nn$ and $pp$ interactions rather than $np$ interaction. Therefore, it is natural that $E_{S,T=0}^{E,\sigma+\omega}$ disappears in recent calculation.

\subsection{Properties of symmetry energy at saturation density}

\begin{table}
\begin{center}
\caption{Kinetic ($\rm{kin}$), isospin-singlet (T=0) and isospin-triplet (T=1) potential components of the symmetry energy $J$ and its slope $L$ at nuclear saturation density with the selected CDF effective interactions. The referred values are taken from BHF calculations in Ref. \cite{IA84}. The values in/out of the parenthesis denote the results with/without the inclusion of the contribution from the Hartree term in $\omega$-meson coupling channel. Units are given in MeV.}
\label{Tab:1}
\setlength{\tabcolsep}{1.5pt}
\begin{tabular}{ccccccr}
\hline\hline
&& TW99 & PKDD  &   PKO1 & PKA1 & Ref. \\ [2.0 ex]
\hline
&kin&8.0&8.1&3.7&0.5&-1.0\\
J&T=0&51.0(8.7)&50.8(10.3)&38.8(11.4)&42.4(10.5)&44.2\\
&T=1&-26.2(16.1)&-22.1(18.4)&-8.1(19.3)&-5.7(26.1)&-9.0\\
\hline
&kin&5.9&5.0&-34.5&-69.6&14.9\\
L&T=0&62.2(-21.1)&78.2(-9.7)&67.5(0.8)&71.3(-14.3)&69.1\\
&T=1&-12.8(70.5)&7.0(94.9)&64.8(131.4)&103.2(188.7)&-17.5\\
\hline\hline
\end{tabular}
\end{center}
\end{table}

It is also interesting to see the properties of symmetry energy at nuclear saturation density. Table.\ref{Tab:1} shows the kinetic ($\rm{kin}$), isospin-singlet (T=0) and isospin-triplet (T=1) potential components of the symmetry energy $J$ and its slope $L$ calculated by the CDF effective interactions. For comparison, we list here as well the results by the microscopic Brueckner-Hartree-Fock approach using the Argonne V18 potential plus the Urbana IX three-body force \cite{IA84}, in which it is exhibited $J$ and $L$ at $\rho_0$ are mainly dominated by T=0 component. Among the CDF results, the RHF effective interactions PKO1 and PKA1 display more accordant values with the referred BHF ones for three components of the symmetry energy $J$, while RMF versions lead to a relatively large negative contribution in T=1 part. Such a redistribution, as has been manifested in Fig. 3, is ascribed to extra contribution from the isoscalar coupling-channels $E_{S,T=1}^{E,\sigma+\omega}$ via the Fock diagram actually. It has been argued that the dominance of T=0 component in symmetry energy is due to the effect of the tensor component of the nuclear force through the $^3S_1-^3D_1$ channel. However, in the CDF calculations the potential via the Hartree diagram plays the role in stead. For slop parameter $L$, although T=0 components in CDF have the similar results to the BHF one, we see that dramatically large uncertainty occurs in kinetic and T=1 parts, where the RMF effective interaction TW99 shows the smallest difference from the BHF one. In accordance, the density-dependence of the symmetry energy at high density becomes quite diverse, and the trend in RHF cases is strongly enhanced.

\begin{table}
\begin{center}
\caption{Contributions from different meson-nucleon coupling-channels to the T=0 and T=1 potential symmetry energy with the selected CDF effective interactions. Units are given in MeV.}
\label{Tab:2}
\setlength{\tabcolsep}{8pt}
\begin{tabular}{cccccc}
\hline\hline
J& channel & TW99 &  PKDD   &   PKO1  &   PKA1  \\ [2.0 ex]
\hline
&$\Gamma_\sigma^D$&47.6&46.2&36.2&38.4\\[0.5 ex]

&$\Gamma_\rho^D$&3.4&4.6&1.5&1.2\\[0.5 ex]

\raisebox{1em}[0pt]{T=0}&$\Gamma_{\sigma+\omega}^E$&0.0&0.0&0.0&0.0\\[0.5 ex]

&$\Gamma_{\rho+\pi}^E$&0.0&0.0&1.2&2.8\\[0.5 ex]
\hline
&$\Gamma_{\sigma}^D$&-36.4&-35.8&-24.6&-23.3\\[0.5 ex]

&$\Gamma_\rho^D$&10.1&13.7&4.4&3.5\\[0.5 ex]

\raisebox{1em}[0pt]{T=1}&$\Gamma_{\sigma+\omega}^E$&0.0&0.0&11.5&12.2\\[0.5 ex]

&$\Gamma_{\rho+\pi}^E$&0.0&0.0&0.6&1.9\\[0.5 ex]
\hline\hline
\end{tabular}
\end{center}
\end{table}

To clarify above discussion we extract further in Table.\ref{Tab:2} the detailed contributions to the T=0 and T=1 potential symmetry energy from the different meson-nucleon coupling-channels with the selected CDF effective interactions. It is observed that in CDF the T=0 potential part of $J$ mainly comes from the Hartree diagram, especially from that in isoscalar $\sigma$-meson coupling-channel, while the T=1 potential contribution to $J$ results from the cancelation between the negative contribution of the Hartree term of $\sigma$-meson coupling-channel and positive contribution of other terms. Since the significant contribution to $J$ from the Fock term of $\sigma-$ and $\omega$-meson coupling-channels in RHF approach, the total T=1 potential symmetry energy is very small for PKO1 and PKA1 effective interactions and is more accordant with the referred BHF calculations.

\begin{table}
\begin{center}
\caption{Similar to Table \ref{Tab:2} but for slop parameter $L$ of symmetry energy.}
\label{Tab:3}
\setlength{\tabcolsep}{8pt}
\begin{tabular}{cccccc}
\hline\hline
L & channel &  TW99 & PKDD & PKO1 & PKA1 \\ [2.0 ex]
\hline
&$\Gamma_\sigma^D$&62.5&69.5&63.8&79.4\\[0.5 ex]

&$\Gamma_\rho^D$&-0.3&8.7&3.7&-0.3\\[0.5 ex]

\raisebox{1em}[0pt]{T=0}&$\Gamma_{\sigma+\omega}^E$&0.0&0.0&0.0&0.0\\[0.5 ex]

&$\Gamma_{\rho+\pi}^E$&0.0&0.0&-0.01&-7.9\\[0.5 ex]
\hline
&$\Gamma_\sigma^D$&-11.9&-19.0&7.5&29.4\\[0.5 ex]

&$\Gamma_\rho^D$&-0.9&26.0&11.1&-0.9\\[0.5 ex]

\raisebox{1em}[0pt]{T=1}&$\Gamma_{\sigma+\omega}^E$&0.0&0.0&44.3&73.4\\[0.5 ex]

&$\Gamma_{\rho+\pi}^E$&0.0&0.0&1.8&1.3\\[0.5 ex]
\hline\hline
\end{tabular}
\end{center}
\end{table}

In Table.\ref{Tab:3} the T=0 and T=1 potential contribution to the slop parameter $L$ is decomposed by the different CDF meson-nucleon coupling-channels. While T=0 component in CDF calculations mainly comes from the Hartree term in the $\sigma$-meson coupling-channel, T=1 component is more complicated, which generates very large uncertainty in sign and magnitude in CDF calculations. It is seen that the inclusion of the Fock terms, in particular by the isoscalar mesons, leads to extra uncertainty of $L$ and thus remarkable difference of the high-density behavior of the nuclear matter.

\section{Summary}
In this paper, the isospin coupling-channels decomposition of the potential energy density functional is performed in the covariant density functional theory, and the isospin-dependence of isospin-singlet (T=0) and isospin-triple (T=1) potential energy is analyzed, i.e., their contributions to the nuclear symmetry energy. We find that the T=0 potential contribution $E_{S,T=0}$ is almost dominated by the Hartree terms in meson-nucleon coupling-channel, while the T=1 potential contribution $E_{S,T=1}$ is obtained by the balance among various coupling channels, especially at high density extra contribution from Fock terms via $\sigma$- and $\omega$-meson coupling-channels. It is shown that in the CDF calculations both T=0 and T=1 potential components play the dominant role in deciding the symmetry energy, especially in RHF cases, which illustrate a quite different mechanism to participate the symmetry energy from the microscopic BHF case, and demonstrate the importance of the Fork diagram (in particular from the isoscalar mesons) in the isospin-related physics.

\section*{Acknowledgements}
The authors thank Bao X. J. for his stimulating discussions. This work is partly supported by the National Natural Science Foundation of China (Grant Nos. 11205075 and 11375076), and the Specialized Research Fund for the Doctoral Program of Higher Education (Grant No. 20120211120002).

\end{document}